\documentclass[12pt,preprint]{aastex}
\usepackage{graphicx,epsfig,wrapfig,paralist}
\usepackage{lscape}
\usepackage{rotating}
\newcommand{\bq}{\begin{equation}}
\newcommand{\eq}{\end{equation}}

\def\gtsim{\lower.5ex\hbox{$\buildrel > \over\sim$}}
\def\ltsim{\lower.5ex\hbox{$\buildrel < \over\sim$}}

\def\apjl{ApJL}
\def\apj{ApJ}
\def\apjs{ApJS}
\def\mnras{MNRAS}
\def\araa{ARAA}

\def\aap{A\&A}
\def\aaps{A\&A Suppl.}
\def\nat{Nature}

\shorttitle{Rotating progenitors of PISNe}
\shortauthors{Chatzopoulos,Wheeler}
\begin{document}
\title
{EFFECTS OF ROTATION ON THE MINIMUM MASS OF PRIMORDIAL PROGENITORS OF PAIR INSTABILITY SUPERNOVAE}
\author{E. Chatzopoulos\altaffilmark{1} \& J. Craig Wheeler\altaffilmark{1}}
%%%  author names
\email{manolis@astro.as.utexas.edu}
\altaffiltext{1}{Department of Astronomy, University of Texas at Austin, Austin, TX, USA.}

\begin{abstract}
 
The issue of which stars may reach the conditions of electron/positron
pair formation instability is of importance to understand
the final evolution both of the first stars and of contemporary stars.
The criterion to enter the pair instability regime in density and
temperature is basically controlled by the mass of the oxygen core.
The main sequence masses that produce a given oxygen core mass are,
in turn, dependent on metallicity, mass loss, and convective and
rotationally-induced mixing. We examine the evolution of massive
stars to determine the minimum main sequence mass that can encounter
pair-instability effects, either a pulsational pair instability (PPISN) or
a full-fledged pair-instability supernova (PISN). We concentrate on 
zero-metallicity stars with no mass loss subject to the Schwarzschild 
criterion for convective instability, but also explore solar metallicity 
and mass loss and the Ledoux criterion. As expected, for sufficiently 
strong rotationally-induced mixing, the minimum main sequence mass is 
encountered for conditions that induce effectively homogeneous
evolution such that the original mass is converted almost entirely
to helium and then to oxygen. For this case, we find that the minimum main sequence mass is 
about 40~$M_{\odot}$ to encounter PPISN and about 65~$M_{\odot}$ to encounter a PISN. 
When mass-loss is taken into account those mass limits become 50~$M_{\odot}$ for PPISN
and 80~$M_{\odot}$ for PISN progenitors. 
The implications of these results for the first stars and for contemporary 
supernovae is discussed.
 
\end{abstract}


\vskip 0.57 in

\section{INTRODUCTION}\label{intro}

Very massive stars were long ago predicted to get hot enough that the ambient photons 
in the interior are sufficiently energetic to create electon/positron pairs 
(Rakavy \& Shaviv 1967; Barkat, Rakavy \& Sack 1967; Rakavy \& Shaviv 1968; 
Rakavy, Shaviv \& Zinamon 1968; 
Fraley 1968; see also Wheeler 1977; El Eid \& Hilf 1977; 
El Eid, Fricke \& Ober 1983; Carr, Bond \& Arnett 1984; Stringfellow \& Woosley 1988).
The conversion of energy to rest mass rather than thermal energy alters the equation 
of state so that the pressure does not increase sufficiently with density upon 
compression to maintain hydrostatic equilibrium. If a sufficiently large, 
mass-averaged region of the star is in the pair-formation regime, such that 
$<\Gamma_{1}>$~$<$~4/3, the structure will be dynamically unstable. Models predict 
that this happens after stars have undergone their central helium burning and have 
formed massive cores composed primarily of oxygen (with a small mass fraction of 
carbon). In models, the instability occurs in an off-center shell. The rapid 
contraction of this shell drives it inward, leading to the rapid compression and 
heating of the inner core of oxygen. Unlike the case of iron-core collapse, the 
oxygen in these stars is subject to strong energy release by rapid thermonuclear 
burning. The result is the prediction that the star is totally disrupted, leaving 
no remnant, but with the production of a very large mass of radioactive $^{56}$Ni, 
the decay of which could power the light output.

The original calculations, cited above, found that the pair-instability regime 
was encountered for stars with massive oxygen cores, greater than about 60~$M_{\odot}$. 
At the time, it was not clear how massive a star needed to be 
to develop a sufficiently massive oxygen core, but estimates were in the range 
of 100~$M_{\odot}$. It was also not clear that any stars sufficiently massive to reach 
this condition of instability existed. The latter point was resolved theoretically 
with the understanding that massive, radiation-pressure dominated stars subject 
to dynamical instability would be stabilized in the non-linear regime (Appenzeller 1970; Ziebarth 1970) 
and observationally with the discovery that young clusters such as R~136 in 30 Doradus contained 
very massive stars (Panagia et al. 1983; Hunter et al. 1995; Crowther et al. 2010). 
A subsequent important development was the prediction that in the context of $\Lambda$CDM models 
of the Universe, the first stars forming at zero metallicity after the Dark Ages 
might preferentially form especially massive stars that would, in turn, be subject 
to pair instability (Abel et al. 1998; Abel et al. 2000; Bromm et al. 2002; Bromm \& Larson 2004). 

Considerable effort has gone into the computation of the formation of the first stars,
the evolution of stars that will reach pair-instability conditions, and the predicted
observational properties of the resulting explosions (Heger \& Woosley 2002, Ohkubo et al. 2003; 
Scannapieco et al. 2005; Blinnikov \& Heger 2007; Waldman 2008; Ohkubo et al. 2009;
Heger \& Woosley 2010; Kasen et al. 2010; 
Whalen \& Fryer 2010; Kasen et al. 2011; Joggerst \& Whalen 2011). 
Pair-Instability Supernovae (PISN) models have a characteristic
nucleosynthetic yield (Heger \& Woosley 2002; Ohkubo et al. 2003), but searches for evidence of such a distribution in
the lowest metallicity stars has not revealed the expected pattern (Christlieb et al. 2002; Frebel et al. 2005). 
In general, lower metallicity will suppress mass loss and allow relatively lower mass main sequence 
stars to encounter the pair-instability regime. For a given main sequence mass, 
higher metallicity will tend to lead to lower mass oxygen cores, thus avoiding this 
regime. Two subsequent developments have altered the context in which pair instability 
is considered. One is the recent understanding that conditions of the first stars 
may be more susceptible to fragmentation so that the first stars may have been of 
smaller mass than once thought (Stacy et al. 2010; Greif et al. 2011).  Another important 
development was the discovery in the contemporary Universe of a category of 
super-luminous supernovae (SLSN) that are relatively rare, but brighter by a 
factor of 10 to 100 than most contemporary supernovae (Smith et al. 2007; Quimby et al. 
2007; Miller et al. 2009; Gal-Yam et al. 2010; Chatzopoulos, Wheeler \& Vinko 2011). 
Some of these SLSN show evidence of high mass, but cannot be 
pair-instability supernovae (Smith et al. 2007; Miller et al. 2009; Chatzopoulos, Wheeler \& Vinko 2011; 
Chatzopoulos, Wheeler \& Vinko 2012; Leloudas et al. 2012). If their 
brightness derived from the radioactive decay of $^{56}$Ni, as demanded by the pair-instability 
model, they would require a greater mass of $^{56}$Ni than is allowed for the total mass of 
the ejecta. The ejecta mass is constrained by the width of the light curve that is a 
measure of the diffusion time. The great luminosity of super-luminous supernovae 
like SN~2006gy probably derives from the collision of the ejecta with a shell of 
matter previously ejected by the progenitor star (Smith et al. 2007; Chevalier \& Irwin 2011;
Chatzopoulos, Wheeler \& Vinko, 2011). Several of these super-luminous 
events do, however, have all the characteristics expected of a PISN (SN 2007bi, Gal-Yam et al. 2010;
PTF 10nmn, Yaron et al. in preparation). These events have occured in low, but not zero, 
metallicity environments. The possibility that the first stars are not of especially
high mass and the discovery of contemporary super-luminous supernovae that have the 
characteristics of PISN brings a new focus to the issue of just which stars can undergo
pair instability and under what conditions.

Maeder (1987; see also Maeder \& Meynet 2011 for a review) discussed the effects of 
rotationally-induced mixing on the evolution of massive 
stars. He concluded that there would be substantial mixing produced by the small-scale 
three-dimensional tail of the turbulent spectrum of the baroclinic instability and that
the diffusion coefficient could be sufficiently large to mix most massive stars during 
their main sequence lifetimes. In particular, above a critical rotation, Maeder predicted
that the evolutionary tracks go upwards and bluewards, very close to those of fully 
homogeneous evolution. The immediate implication was that stars in the mass range predicted
to reach pair instability (already radiation-pressure dominated and hence close
to neutral dynamical instability) were susceptible to quasi-homogeneous evolution such
that nearly all the main sequence mass could be burned to heavier elements. Quasi-homogeneous evolution would 
significantly decrease the main sequence mass that leads to pair instability. 
The notion that quasi-homogeneous evolution might lead to larger core masses has been applied in the 
context of the progenitors of GRBs (Heger, Woosley \& Spruit 2005; Yoon \& Langer 2005) and explored to a 
certain extent in recent work (Langer et al. 2007, Eskstrom et al. 2008, 2011; Brott et al. 2011a, 2011b), 
but has not been pursued in detail in the context of pair instability. 

In this paper, we conduct a thorough parameter study to explore the minimum main sequence
mass that encounters the pair instability regime. In his unpublished PhD thesis, Barkat (1967)
noted that some models that skirted the pair-instability regime ejected shells of matter, but
survived the instability and continued to evolve. Somewhat higher mass models became sufficiently
unstable that a dynamical explosion, a PISN, was produced directly. Woosley, Blinnikov \& Heger (2007) 
invoked the former effect, that they called a Pulsational Pair Instability Supernova (PPISN) to address 
the nature of SLSN~2006gy. Their simulations showed a repeated ejection of shells. The second, 
faster shell collided with the first and produced a luminous display reminiscent of the light 
curve of SLSN~2006gy. Here we will delineate models that encounter the PPISN and those that encounter
the full PISN explosion. Section 2 describes our assumptions and models, \S 3 gives the
results and presents a model that invokes a different mixing criterion and another
with solar metallicity and mass loss. Section 4 discusses our conclusions. 

\section{MODELS}\label{mods}

As discussed in the Introduction, the masses of stars that evolve from the main sequence
to either PPISN or PISN will be a function of metallicity, mass loss, and rotationally-induced mixing.  
The outcome will also depend on the treatment of convective instability, semi-convection,
and overshoot. Dissipation of shear by magnetic effects may also attend the rotationally-induced
mixing. Rather than explore this whole parameter space where the physics is, in any case, 
uncertain, we have focused on the portion of parameter space that is expected to lead to 
the minimum mass to encounter pair formation. In particular, we have explored conditions of
zero metallicity and have neglected mass loss for the majority of our models in order
to establish the proof of principle. We have run a few models at solar metallicity in order
to put our results in context. We have adopted rates of rotation on the main sequence
that run from non-rotating to rotating at 80 \% of equatorial Keplerian velocity. We have
focused on Schwarzschild convection for two reasons. One is that this prescription will tend
to enhance the mass of the oxygen core for a given main sequence mass, in keeping with the
philosophy of this exploratory work. The second reason is based on a suspicion that neither
the Schwarzschild nor the Ledoux criterion really captures the three-dimensional, plume-driven
convection in real stars. As we will describe in a future work (Chatzopoulos, Dearborn \&
Wheeler, in preparation), we have grounds to believe that in the late stages, when the 
oxygen core forms, the Schwarzschild criterion is the more appropriate. Multi-dimensional
effects tend to swamp the stabilizing effect of composition gradients. Our simulations have
also adopted the effects of magnetic viscosity as parametrized by Heger, Woosley \& Spruit (2005) based
on the prescriptions of Spruit (1999, 2002).
This is not because we believe that this particular parametrization captures all the relevant
multi-dimensional MHD instabilities and related phenomona in rotating, shearing, stars, but
because it is a widely used and recognized algorithm so that our results can be readily
compared to others using the same prescription.

We have used the Modules for Stellar Experiments in Astrophysics (MESA; Paxton et al. 2011) code to calculate the evolution
of a grid of $Z =$~0 massive stars ranging from Zero Age Main Sequence (ZAMS) mass of 35~$M_{\odot}$ to 200~$M_{\odot}$ for four
different degrees of ZAMS rotation. We assume initially rigid body rotation on the ZAMS with surface rotation corresponding to
0, 30\%, 50\% and 80\% of the critical Keplerian rotation $\Omega_{crit} =$~$(g(1-\Gamma)/R)^{1/2}$ where
$g =$~$GM/R^{2}$ is the gravitational acceleration at the ``surface" of the star, $G$ the gravitational constant, $M$ the
mass, $R$ the radius of the star and $\Gamma=L/L_{Ed}$ the Eddington factor where $L$ and $L_{Ed}$ is the total radiated
luminosity and the Eddington luminosity respectively. 
The 35~$M_{\odot}$ model was only run for maximum rotation (80\%) in order to establish
the lower mass limit for PPISN and the models above 85~$M_{\odot}$ for lower (30\%) and zero degrees of rotation to establish
the corresponding minimum ZAMS mass for PPISN. The models with ZAMS masses 40-85~$M_{\odot}$ were run in bins of 5~$M_{\odot}$ 
for all the selected degrees of rotation in order to better resolve the limits for different final fates of the stars.
In order to benchmark against the results of Heger \& Woosley (2002) and Woosley, Blinnikov \& Heger (2007) for non-rotating stars
that will produce a pure PISN and a PPISN, respectively, we have run a 110~$M_{\odot}$ and a 200~$M_{\odot}$ model (ZAMS masses) 
without rotation. The reason that we chose this range of ZAMS masses is the fact that Heger \& Woosley (2002)
and Heger et al. (2003) predict that oxygen core masses ($M_{O-core}$) ranging from 40-64~$M_{\odot}$ will undergo PPISN while 
64~$< M_{O-core} <$~133~$M_{\odot}$ will explode as direct PISNe for non-rotating progenitors,
but with significantly rapid rotation a lower ZAMS mass can produce oxygen core masses in this range. 

MESA was run with the Schwarzschild criterion for convection implemented for reasons discussed above and for zero mass loss. It should
be mentioned that rotationally induced mass loss may reduce the final oxygen core masses. Ekstrom et al. (2008) found that rapidly 
rotating zero metallicity stars with ZAMS masses above $\sim$~50~$M_{\odot}$ may lose up to $\sim$~11~$M_{\odot}$; however due to
the lack of understanding of mass loss mechanisms in massive stars this remains uncertain. Nevertheless,
we also conducted calculations with mass loss included in models with initially zero metallicity, as described in \S3.2, to estimate
the impact of mass loss on the final fate of the models.
MESA employs a combination of prescriptions
for the equation of state (EOS), but for high density and temperature plasma the HELM EOS (Timmes \& Swesty 2000) is used. The HELM EOS
accounts for pressure induced by radiation, ions, electrons, positrons and corrections for Coulomb effects and therefore for the effects
of electron-positron pair formation, which drives the adiabatic index $\Gamma_{1}$ below 4/3. For the treatment of nuclear processes 
with MESA we employ the ``approx21" network (Timmes 1999), which covers all major stellar nuclear reaction rates. 
The effects of angular momentum transport via rotation and magnetic fields are treated based on the one-dimensional approximations of 
Spruit (1999, 2002) and Heger, Woosley \& Spruit (2005).

MESA is capable of running stellar models up to the core collapse (CC) and pre-supernova (pre-SN) stage;
however for high mass stellar models
that encounter a degree of instability induced by electron-positron pair production the effects become very dynamic  and a challenge
for a stellar
evolution code to handle. Future expansions of MESA will be able to handle those dynamical effects and to follow the supernova (SN)
explosions of stars with 1-D hydrodynamics implemented (Paxton; private communication). In the current work, 
stellar models for which a signficant
fraction of their core approaches the $\Gamma_{1} <$~4/3 regime due to pair formation are stopped before core oxygen ignition and
within the carbon burning phase, shortly before becoming dynamic. Those models are then mapped into the 
multi-dimensional hydrodynamics code FLASH (Fryxell et al. 2000), and their evolution is followed in 1-D. The newest version of FLASH
(FLASH4-alpha release) is used for these simulations. FLASH is very suitable to follow the dynamical 
transition of the models from MESA because
it uses the same EOS (HELM; Timmes \& Swesty 2000) and similar nuclear reaction network. 
PPISNe are characterized by a violent contraction and pulsation that heats the core up to a temperature such that some of the oxygen
is burned to produce primarily $^{28}$Si and $^{32}$S (Woosley, Blinnikov \& Heger 2007). Pure thermonuclear PISNe are heated 
significantly enough from the dynamical collapse induced by the softening of the EOS due to electron-positron pair formation that they
burn oxygen explosively and large amounts of $^{56}$Ni are produced. FLASH is capable of reproducing those basic features of the 
events and is therefore used to establish the final fate of the models. We note that in the 1-D FLASH hydrodynamic simulations
the effects of rotation are not considered. In a future work we plan to investigate the effects of rotation in the hydrodynamical
stage of PISNe with multi-dimensional FLASH simulations. 

\section{RESULTS}\label{results}

Table 1 summarizes the characteristics of all the models considered in this work (for zero metallicity and mass loss turned off). 
The first column lists the ZAMS mass ($M_{ZAMS}$,
in solar masses), 
the second is the critical rotational ratio $\Omega/\Omega_{crit}$, the third and fourth the maximum central temperature (in units of
$10^{9}$~K and density (in units of $10^{5}$~g~cm$^{-3}$) that were encountered due to the pair formation dynamical instability,
the fifth the mass of the oxygen core, $M_{O-core}$ (in units of solar masses), 
that each model produced, the sixth the surface abundance of $^{14}$N and
the seventh the final fate of the model as observed in the FLASH hydrodynamics simulations (CC for core collapse, PPISN for pulsational
pair instability supernova and PISN for pair instability supernova). In all figures, the models that ended as CC will
be represented by black curves, the models that encountered PPISN by green curves and the models that underwent PISN with red curves.

To compare with previous results for PPISN presented by Woosley, Blinnikov \& Heger (2007) in the case of a 110~$M_{\odot}$ ZAMS star
and for direct PISN presented by Heger \& Woosley (2002) for a variety of stars with $M_{ZAMS} >$~140~$M_{\odot}$ we run our own
non-rotating 110~$M_{\odot}$ and 200~$M_{\odot}$ models. The model with 110~$M_{\odot}$ formed a 56~$M_{\odot}$ oxygen core and then
encountered a violent pulsation that heated the center of the star up to $2.46 \times 10^{9}$~K, in a similar manner to that 
suggested by Woosley, Blinnikov
\& Heger (2007). The 200~$M_{\odot}$ model produced a direct PISN that synthesized a massive amount of $^{56}$Ni
($\sim$~21~$M_{\odot}$) and totally disrupted the star, behaving exactly as predicted by Woosley, Blinnikov \& Heger (2007). 
Additionally, in order to establish a lower mass end for the production of PPISNe, we have run a 35~$M_{\odot}$ model with rotation at 80\% 
of the critical velocity that converted essentially 
all of its mass to oxygen. This model was able to evolve up to CC in MESA. For benchmarking
it was also mapped to FLASH at the time of core oxygen ignition where it also kept slowly evolving toward higher densities and temperatures
always avoiding the pair formation regime. 

For all the models, increased levels of rotation led to a more chemically homogeneous evolution and
produced higher oxygen core masses. ZAMS rotation at 80\% of the critical
Keplerian velocity was able to convert all stellar mass into oxygen for all the models. 
The evolution of some models rotating at 80\% yielded brief stages when rotation became mildly super-critical (110-120\%) 
at the surface which means that results from these models may not be completely accurate. For all other degrees of rotation the star
remained at sub-critical velocities throughout all of its evolutionary track. As a representative example of this effect, we show 
the evolution of $\Omega/\Omega_{crit}$ for the rotating 70~$M_{\odot}$ models during the main-sequence (MS) in Figure 1. The solid green
curve, the dashed green curve and the solid red curve show the evolution of $\Omega/\Omega_{crit}$ for ZAMS rotation at 30, 50 and 80\%,
respectively. 

In order to illustrate the effects of rotation on a ZAMS star with a specific mass we pick the 70~$M_{\odot}$ models because, as can
be seen in Table 1, all possible final fates are encountered for this model (CC for zero rotation, PPISN for 30\% and 50\% critical
rotation and PISN for 80\% critical rotation). The evolution of the central density ($\rho_{c}$) and temperature ($T_{c}$) for
these models is presented in the left panel of Figure 2. As can be seen, all rotating models encounter a collapse that heats the core
up to higher densities and temperatures, but the one rotating at 80\% critical is heated significantly enough to burn oxygen explosively
and subsequently ejects of all its mass. Higher degrees of rotation lead the $\rho_{c}$-$T_{c}$ track of the star closer to the 
$\Gamma_{1} <$~4/3 pair formation region. The right panel of Figure 2 shows the tracks of the models in the Hertzprung-Russell diagram.
It can be seen that the more rapidly rotating models remain bluer and are more luminous that the less rapidly rotating ones. 
Figure 3 illustrates the chemical composition of the 70~$M_{\odot}$ models for all degrees of rotation at the time before core
oxygen ignition (upper left panel: no rotation;
upper right panel: 30\% critical rotation; lower left panel: 50\% critical rotation; lower right panel: 80\% critical rotation). The 
effects of increased homogeneity and higher $M_{O-core}$, with increasing rotation are clearly illustrated. 
Figure 4 presents the distribution of the CNO (solid
black curve) and 3-$\alpha$ process specific nuclear energy inputs are shown. 

Higher degrees of rotation also produced a significantly higher $^{14}$N surface abundance than their non-rotating 
counterparts, a trend also noted by Ekstrom et al. (2008). 
This trend, however, is not monotonic. For the extreme level of 80\% critical rotation the surface $^{14}$N mass fraction is  generally
found to be reduced while the most significant $^{14}$N enrichment is observed in models rotating at 50\% of the critical
value. Increased surface $^{14}$N is attributed to the onset of the CNO cycle in the outer shell due to the strong rotationally
induced mixing. For completely homogeneous evolution that makes an oxygen star, which takes place for models that rotate
at the extreme level of 80\% critical rotation, the CNO contribution in outer layers of the star is confined to a very thin shell. 

Figure 5 illustrates our final results for the fate of rotating massive primordial stars. As can be seen,
stars initially rotating with speeds 80\% the critical velocity with 40~$M_{\odot}$~$< M_{ZAMS} <$~60~$M_{\odot}$ will produce PPISNe
associated with episodic mass-loss and for $M_{ZAMS} \geq$~65$~M_{\odot}$ they will explode as PISNe. As mentioned before,
at this rapid rotation the whole mass of those stars will turn into oxygen, therefore this result is consistent with the findings
of Heger et al. (2003), but for initial ZAMS masses that are only 40-50\% those of the ones given by their non-rotating calculations.
Consequently, for this ``fiducial" degree of rotation at 50\% the critical value, the ZAMS mass limits become 
45~$M_{\odot}$~$< M_{ZAMS} <$~70$~M_{\odot}$ for PPISN and $M_{ZAMS} \geq$~75~$M_{\odot}$ for PISN progenitors, thus at the level of
50-60\% of those in the case of no rotation. 
These results suggest that episodic mass-loss events resulting from PPISNe can be
encountered for less massive stars and may account for some of the observed LBV-type events. We wish to add that a recent paper
by Yoon, Dierks \& Langer (2012) independently calculated the mass limits of PPISN and PISN primordial progenitors and found good
agreement with our results. 

\subsection{{\it Effects of Mixing}}\label{mixeff}

Convective mixing and overshoot are some of the factors that control the final mass of the oxygen core. As mentioned above, in this
work we have adopted the Schwarzschild criterion for convection based on the lack of true knowledge about the nature of convection
in realistic three-dimensional situations. One dimensional convective mixing will be supressed if composition gradients are considered,
as suggested by the Ledoux criterion for convection. This will result in smaller final oxygen core masses and therefore tracks in
the $\rho_{c}$-$T_{c}$ plane that are shifted further away from the pair formation $\Gamma_{1} <$~4/3 region, something that may
alter the results for the minimum ZAMS masses that encounter PPISN and PISN. 

We ran MESA for the 70~$M_{\odot}$ non-rotating and 50\% critically rotating models with the Ledoux criterion implemented in order
to examine the sign of this effect. Figure 6 presents the results in the case of zero rotation.
The left panel shows the $\rho_{c}$-$T_{c}$ tracks
for the model that uses the Schwarzschild (solid black curve) and the model that uses the Ledoux (dashed black curve) criterion. The right
upper and lower panels show the chemical composition of the two models at the time before core oxygen ignition. 
It can be seen that when mixing is supressed the 
non-rotating 70~$M_{\odot}$ model evolves towards slightly lower $\rho_{c}$ and $T_{c}$ values and it ends up making an oxygen
core that is 86\% the one produced in the case where composition gradients are ignored. 
The result of the same experiment but in the case of rotation at the 50\% of the critical value is shown in Figure 7. In this
case, we see that the fast rotationally-induced mixing counters the effects of suppression due to the inclusion of composition gradients
and the final oxygen core masses are almost equal. The characteristics of the models run with the Ledoux criterion implemented
are also given in Table 1. These results indicate that selecting a different one-dimensional prescription
for convective mixing is unlikely to alter our results for the rotating models by any significant factor.

\subsection{{\it Effects of Metallicity and Mass Loss}}\label{zeff}

Although a thorough study of massive models with $Z >$~0 is beyond the scope of this project, which focuses
on the primordial progenitors of PISNe, we considered the effects that the presence of metals may have.
As is well known, metallicities with $Z >$0 can induce significant line-driven mass loss that can cause massive stars
to lose a significant amount of mass. This mass loss will drive evolution towards the formation of oxygen cores with smaller mass
than in the case
of zero mass loss. This will have an effect on the final fate of a very massive star, such that it might miss the pair
formation region in the $\rho_{c}$-$T_{c}$ plane and end its life as a CC SN explosion. The high mass loss may lead
to the formation of massive circumstellar material (CSM) environments around the progenitors of these SN
so that the SN ejecta will violently interact with the CSM producing shock energy that can power their light curve (LC), 
as manifested by luminous Type IIn SNe. 
 
To account for the mass loss with MESA, the prescriptions of de Jager, Nieuwenhuijzen \& van der Hucht (1988)
were used, as appropriate for hot O-type stars (for a recent discussion on mass loss rates for Wolf-Rayet stars, see
Yoon, Woosley \& Langer 2010). 
Figure 5 also shows the final fate of the same rotating massive zero metallicity stars for which mass loss is considered in the
calculations. As mentioned above, rotationally-induced mixing will dredge up metals to the outer and less gravitationally bound
regions of the star inducing line-driven mass loss that will ultimately result in smaller oxygen core mass. 
We find that when mass loss is considered, the minimum ZAMS mass that produce PPISN events is $\sim$~50~$M_{\odot}$ 
and the minimum mass that produce PISN events is $\sim$~80~$M_{\odot}$ at maximum ZAMS rotation (80\% of the critical value).
The results for the fate of the 70~$M_{\odot}$ model rotating at 50\% the critical value with solar metallicity and mass loss
are presented in Figure 8.
The left panel of Figure 8 shows a comparison
between the $\rho_{c}$-$T_{c}$ tracks of the two models with (solid black curve) and without (solid green curve) mass loss, 
and the upper and lower right panels show the chemical composition
of the models at the time before core oxygen ignition. The strong line-driven mass loss ejected the hydrogen and a fraction
of the helium shell leaving a 33~$M_{\odot}$ star with $M_{O-core} =$~16~$M_{\odot}$ that ended its life as a CC SN. The
strong continuous mass loss led the $\rho_{c}$-$T_{c}$ track of the model further from the electron-positron pair $\Gamma_{1} <$~4/3
region. Furthermore, due to strong rotationally-induced mixing the pre-SN model had an enhanced $^{14}$N surface abundance as was the
case with most rotating $Z =$~0 models. The extreme amount of mass lost by the star that we compute here is consistent with
results presented by Ekstrom et al. (2011) who calculated the evolution of a grid of solar metallicity models 
from 0.8 to 120~$M_{\odot}$ rotating at 40\% the critical velocity. Ekstrom et al. (2011) found that 
even the most massive rotating 120~$M_{\odot}$ model
ends its life with a mass of 19~$M_{\odot}$, way below the limit for PPISN. This result indicates that PPISN and, even more so, 
PISN events that may result from local ($Z \sim$~$Z_{\odot}$) massive progenitors must be very rare compared to those that result
from metal-poor primordial progenitors. 

\section{DISCUSSION AND CONCLUSIONS}\label{disc}

Recently, Stacy et al. (2011) determined the rotational speed of massive primordial stars to be close to 50\% of the critical value
using multi-dimensional smooth particle hydrodynamics (SPH) simulations. In addition, observational evidence presented by
Dufton et al. (2011) on the 20-30$M_{\odot}$ late O-type star VFTS102 in 30 Doradus indicates a rotational speed
$>$~50\% of the critical value for this star. VFTS102 has $Z >$~0 and is not massive enough to encounter pair formation in 
any circumstances, but it establishes the existence of fast rotation in some high mass stars.
Using MESA to follow the evolution and FLASH to compute the late time hydrodynamics for a grid of massive primordial ($Z =$~0) stars
with various degrees of rotation and including the effects of magnetic fields as currently parametrized,
we established the ZAMS mass ranges required to produce PPISN or PISN as presented in Figure 5.
For significant rotational velocities, we find that stars with ZAMS masses as low as 40-45~$M_{\odot}$ can produce PPISN
and stars with ZAMS masses $>$~65-75~$M_{\odot}$ can produce direct PISN explosions due to a more chemically homogeneous evolution
that leads to increased oxygen core masses.

We also investigated the effects of convective mixing, metallicity and mass loss. We found that suppressed mixing
due to composition gradients is unlikely to significantly alter the results for the rotating models since
rotationally-induced mixing is the dominant factor. We also found that higher metallicities (and specifically solar metalicity)
can induce extreme line-driven mass loss so that even some of the most massive ZAMS stars end their lives as CC SNe avoiding
the pair-formation instability. 
Current results from numerical simulations (Greif et al. 2011) suggest a nearly flat primordial initial mass function 
(IMF) with typical mass of 100~$M_{\odot}$. If so, PISN and especially PPISN events may be more frequent than 
estimates suggest based on non-rotating models (Scannapieco et al. 2005). This raises the possibility of detecting
a larger number of those spectacular explosions with future missions such the James Webb Space Telescope (JWST).

As emphasized by Smith \& Owocki (2006) and Smith et al. (2007), stars under a
wide variety of conditions may undergo the mass ejection process 
associated with luminous blue variables (LBV). The physical mechanism of
the LBV phenomenon is not well understood. PPISN is not likely to be the
only mechanism involved in the LBV process, but our result that stars with
initial mass as low as 40~$M_{\odot}$ may undergo PPISN means that PPISN should be
considered as a possible candidate mechanism for the LBV mass loss 
phenomenon in some circumstances. It would be worthwhile to explore the 
PPISN process in stars with a variety of envelope compositions and structures, 
including those that are nearly pure, bare, oxygen cores to understand how 
PPISN may lead to single or multiple shell-ejection phases.

We thank the MESA team for making this valuable tool readily available
and especially thank Bill Paxton for his ready advice and council 
in running the code. We wish to thank the anonymous referee for useful suggestions
and comments.
We also wish to thank Volker Bromm and Athena Stacy for discussions
of the formation and evolution of Pop III stars. EC would like to thank Tuguldur Sukhbold for
help with MESA and Selma de Mink for useful suggestions. EC would also like to thank Sean Couch and
Christopher Lindner for help with FLASH. JCW
especially thanks Zalman Barkat for first introducing him to this
topic and for sharing portions of his PhD thesis. This research is supported in part by NSF AST-1109801.

%%%%%%%%%%%%%%%%%%%%%%%%%%%%%%%%%%%%%%%%%%%%%%%%%%%%%%%%%%%%%%%%%%%%%%%%%%%%
%%            REFERENCES
%%%%%%%%%%%%%%%%%%%%%%%%%%%%%%%%%%%%%%%%%%%%%%%%%%%%%%%%%%%%%%%%%%%%%%%%%%%%

{}                     
%\end{references}

%%%%%%%%%%%%%%%%%%%%%%%%%%%%%%%%%%%%%%%%%%%%%%%%%%%%%%%%%%%%%%%%%%%%%%%%%%%%
%%            FIGURES 

\begin{figure}
\begin{center}
\includegraphics[angle=-90,width=13cm]{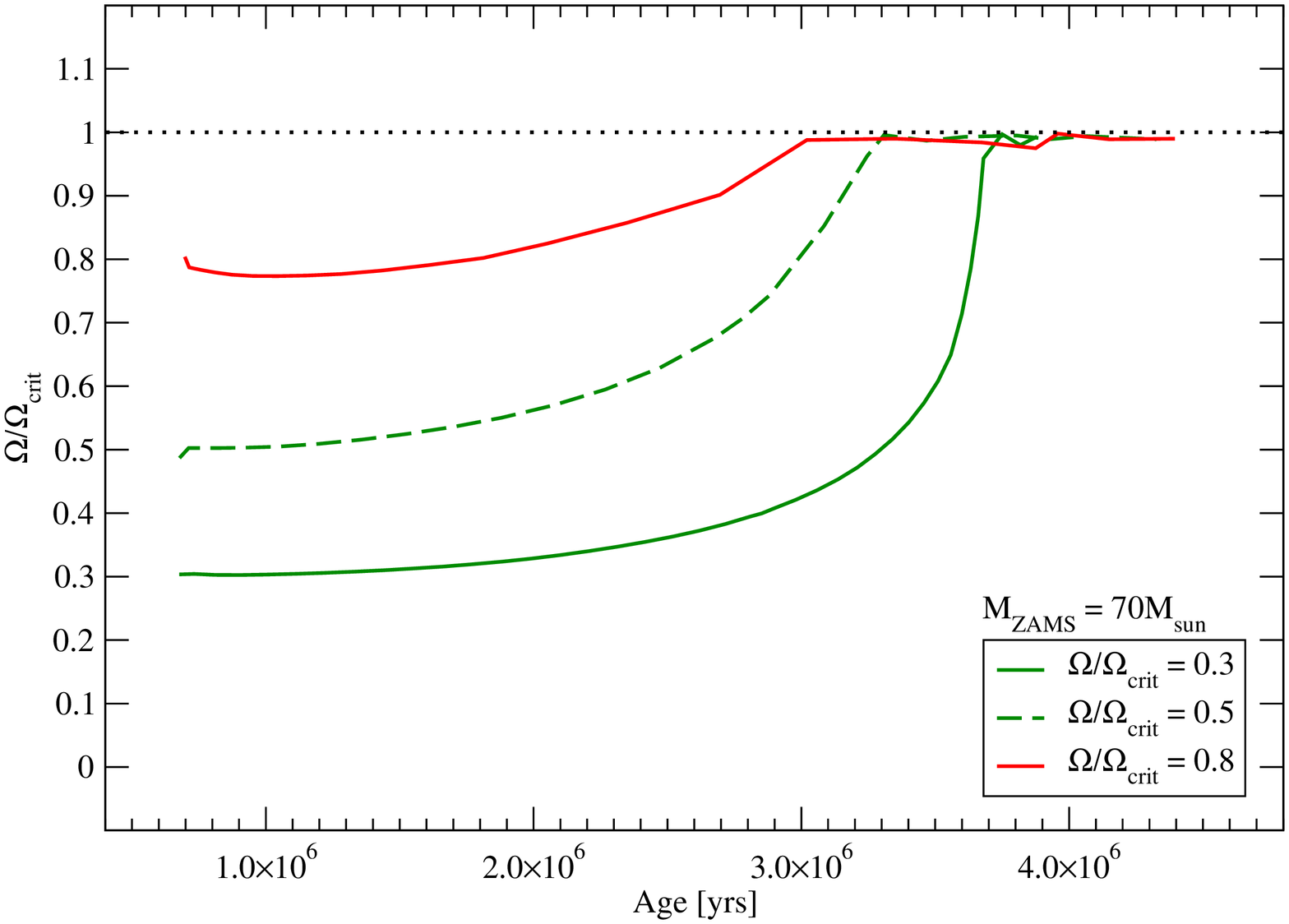}
\caption{Evolution of $\Omega/\Omega_{crit}$
at the surface for the 70~$M_{\odot}$ models for initial ZAMS
$\Omega/\Omega_{crit} =$~0.3 (solid green curve), $\Omega/\Omega_{crit} =$~0.5 (dashed green curve)
and $\Omega/\Omega_{crit} =$~0.8 (solid red curve).}
\end{center}
\end{figure}

\begin{figure}       
\centerline{
\hskip -0.2 in
\psfig{figure=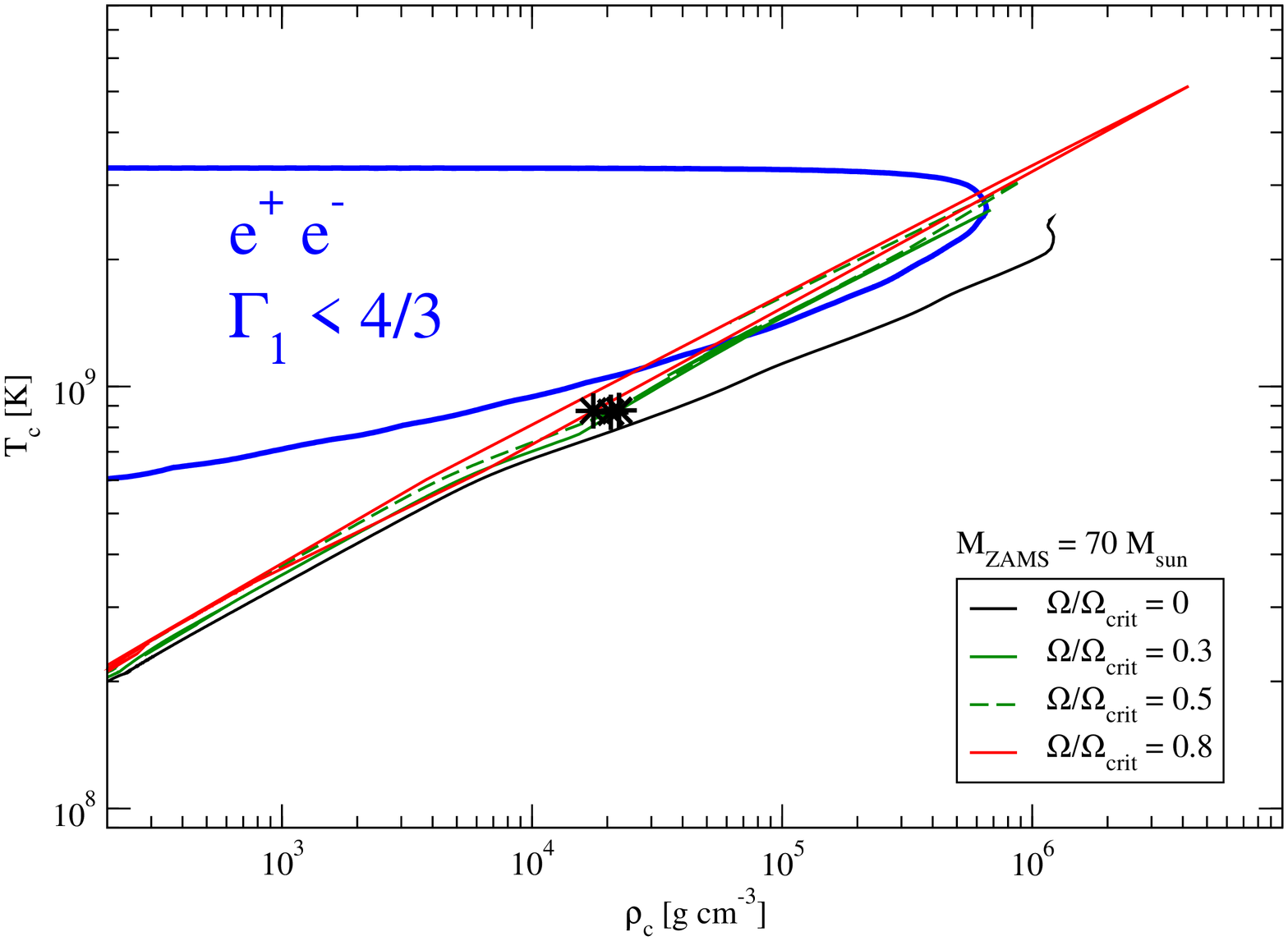,angle=-90,width=3in}
%\hskip 3.0 in
\psfig{figure=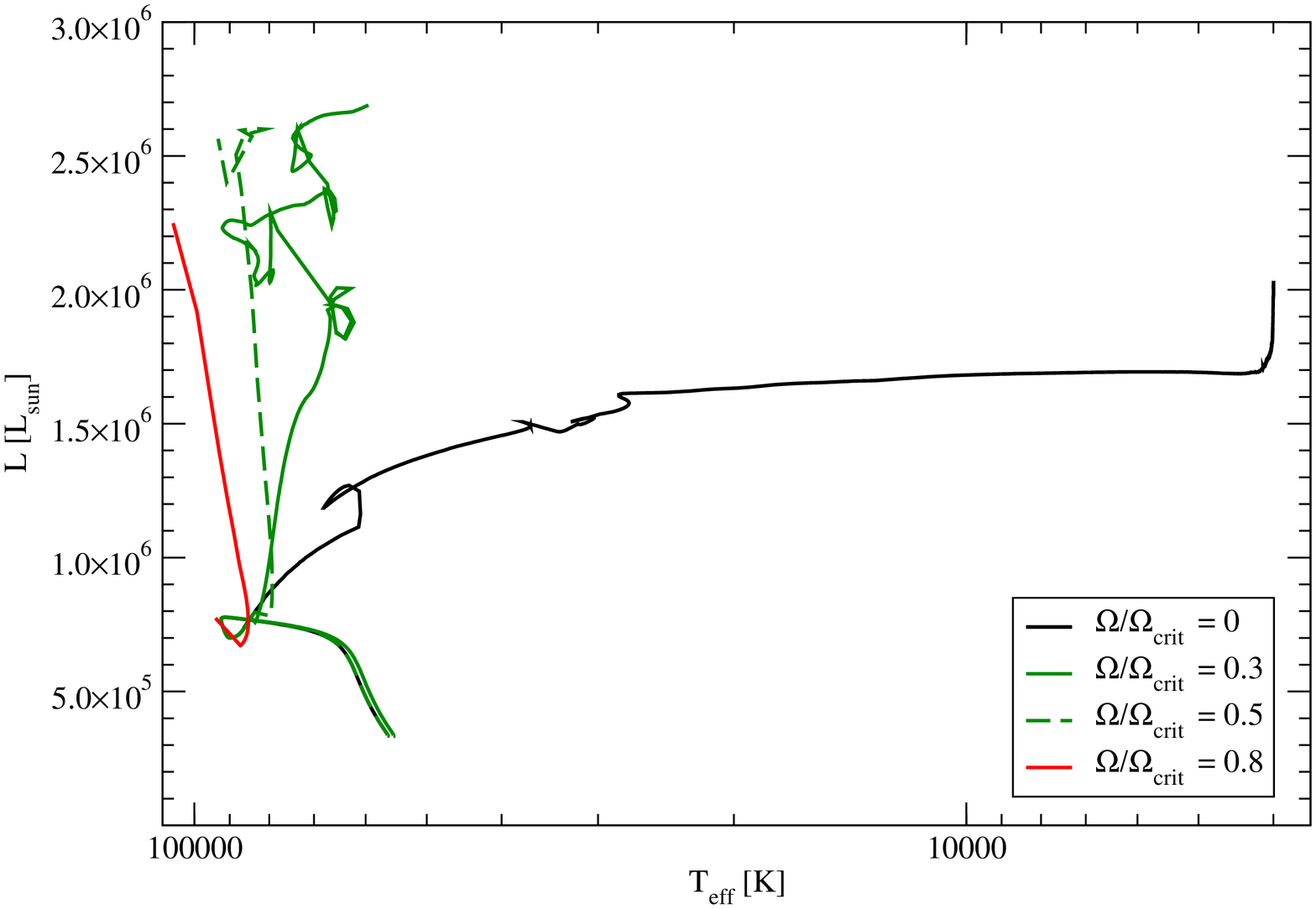,angle=-90,width=3in}
}
%\vspace {-60 pt}
\caption{{\it Left Panel}: Central density and temperature 
evolution of the 70~$M_{\odot}$ (Schwarzschild)
models for $\Omega/\Omega_{crit} =$~0 (solid black curve),
$\Omega/\Omega_{crit} =$~0.3 (solid green curve),
$\Omega/\Omega_{crit} =$~0.5 (dashed green curve), and
$\Omega/\Omega_{crit} =$~0.8 (solid red curve). The solid blue curve 
marks the electron-positron pair instability region where the adiabatic
index is $\Gamma_{1} < 4/3$. The black stars mark the point where the models were mapped to
the hydrodynamics code. 
{\it Right Panel}: Evolution of the 70~$M_{\odot}$ (Schwarzschild) models in the H-R diagram.}
\end{figure}

\begin{figure}
\begin{center}
\includegraphics[angle=-90,width=18cm]{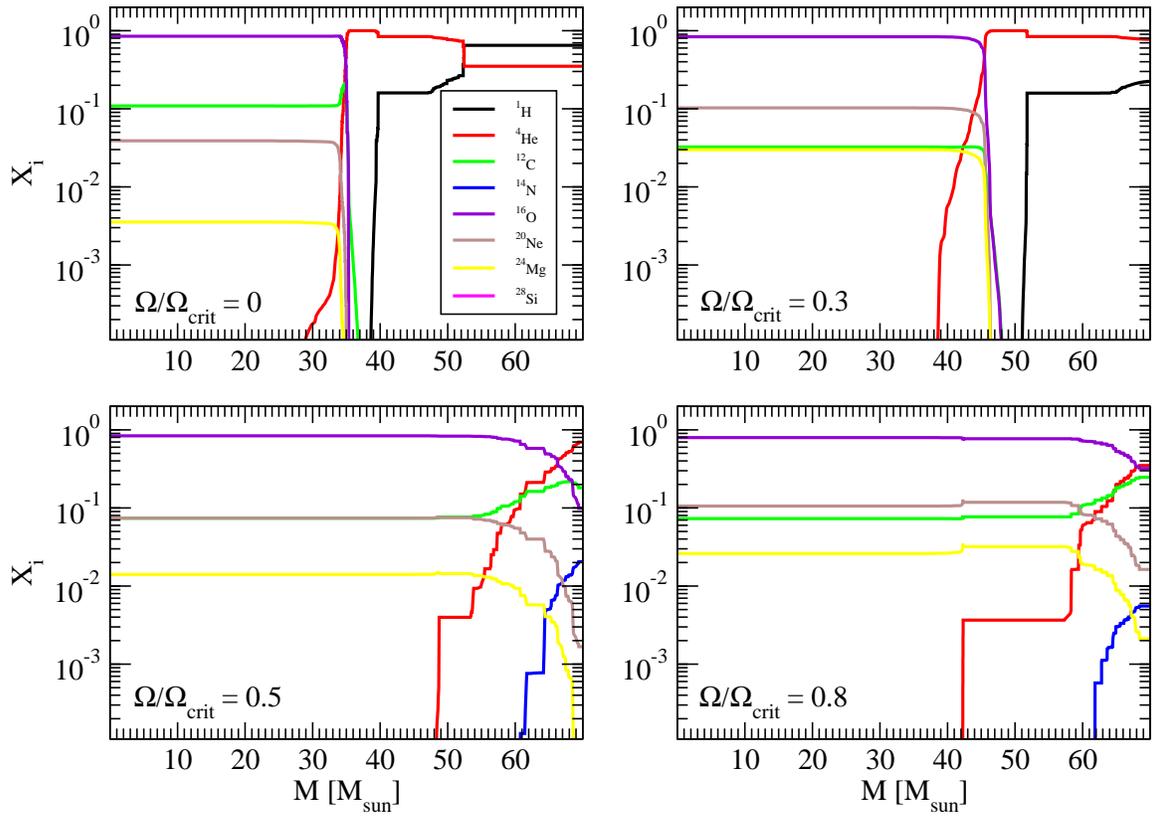}
\caption{Chemical composition of the 70~$M_{\odot}$ (Schwarzschild) models
for $\Omega/\Omega_{crit} =$~0 (upper left panel),
$\Omega/\Omega_{crit} =$~0.3 (upper right panel),
$\Omega/\Omega_{crit} =$~0.5 (lower left panel), and
$\Omega/\Omega_{crit} =$~0.8 (lower right panel) 
at the time just prior to core oxygen ignition.
The specific elements plotted are
given in the inset in the upper left panel.}
\end{center}
\end{figure}

\begin{figure}
\begin{center}
\includegraphics[angle=-90,width=18cm]{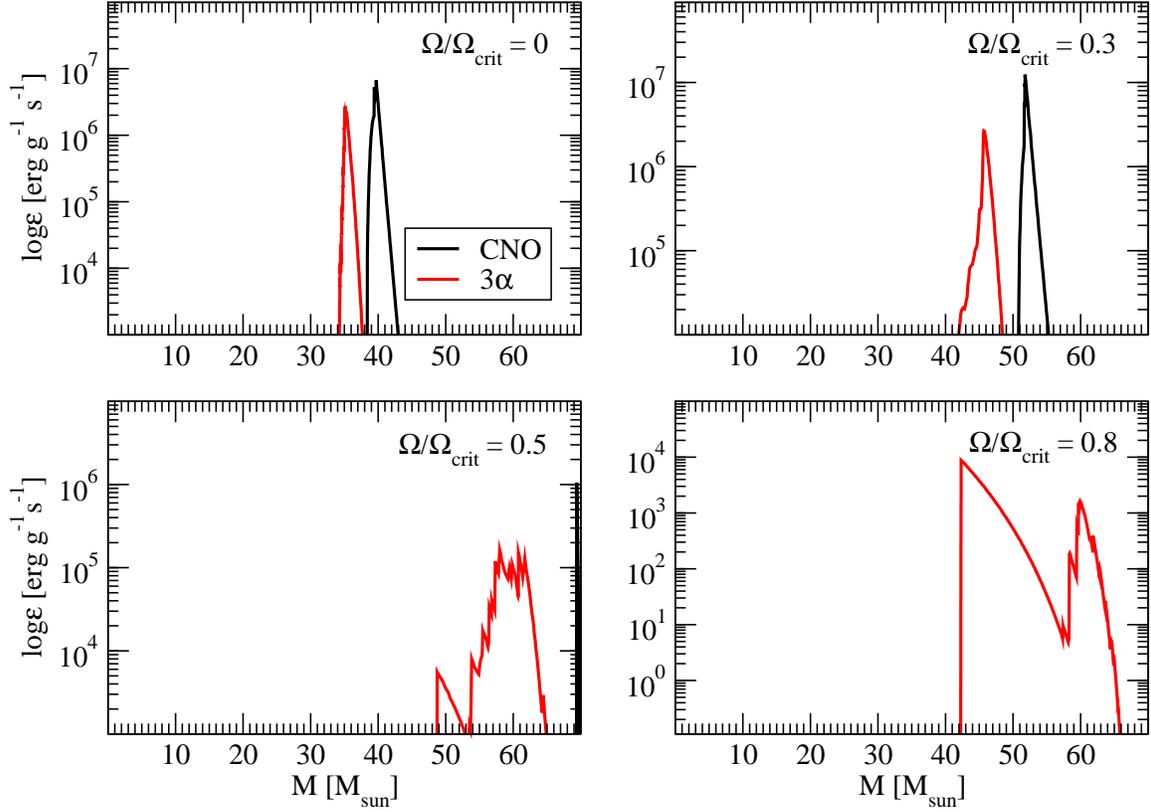}
\caption{Specific nuclear energy generation input mass distributions 
for the 70~$M_{\odot}$ (Schwarzschild) models
(CNO cycle: solid black curve; triple-alpha process: solid
red curve) for  $\Omega/\Omega_{crit} =$~0 (upper left panel),
$\Omega/\Omega_{crit} =$~0.3 (upper right panel),
$\Omega/\Omega_{crit} =$~0.5 (lower left panel), and
$\Omega/\Omega_{crit} =$~0.8 (lower right panel)
at the time just prior to core oxygen ignition.}
\end{center}
\end{figure}

\begin{figure}
\begin{center}
\includegraphics[angle=-90,width=18cm]{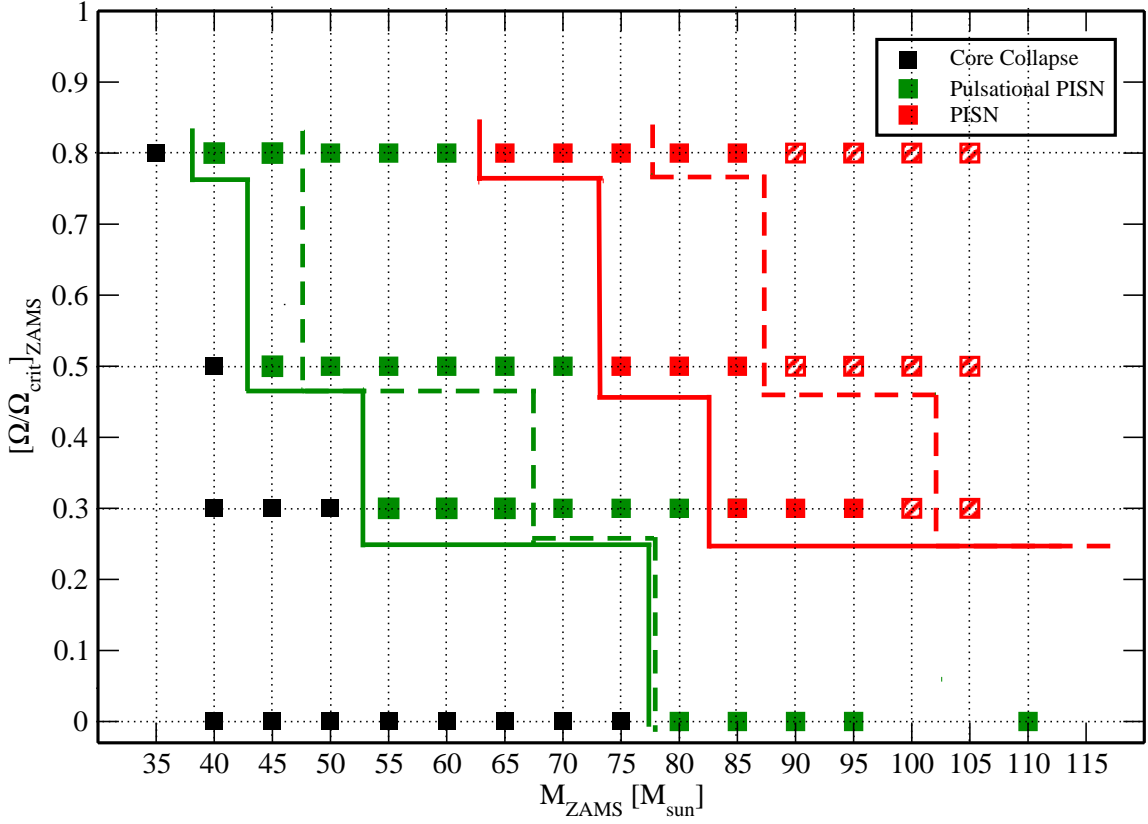}
\caption{Final fate of high ZAMS mass stellar models for different degrees of ZAMS rotation. Filled black
squares indicate core collapse (CC), filled green squares pulsational pair instability (PPISN) and filled red
squares direct pair instability supernova explosion (PISN). 
The hatched red squares indicate the PISN fate of the $Z =$~0 models run only with mass-loss included in the calculations.
The dotted lines mark the ZAMS mass and rotation level
grid for the models done in this work. The thick green and red lines mark the approximate boundaries between the different
final fates of the models. The dashed thick green and red lines mark those same approximate boundaries in the case
where mass-loss at $Z =$~0 is considered in the calculations.}
\end{center}
\end{figure}

\begin{figure}
\begin{center}
\includegraphics[angle=-90,width=18cm]{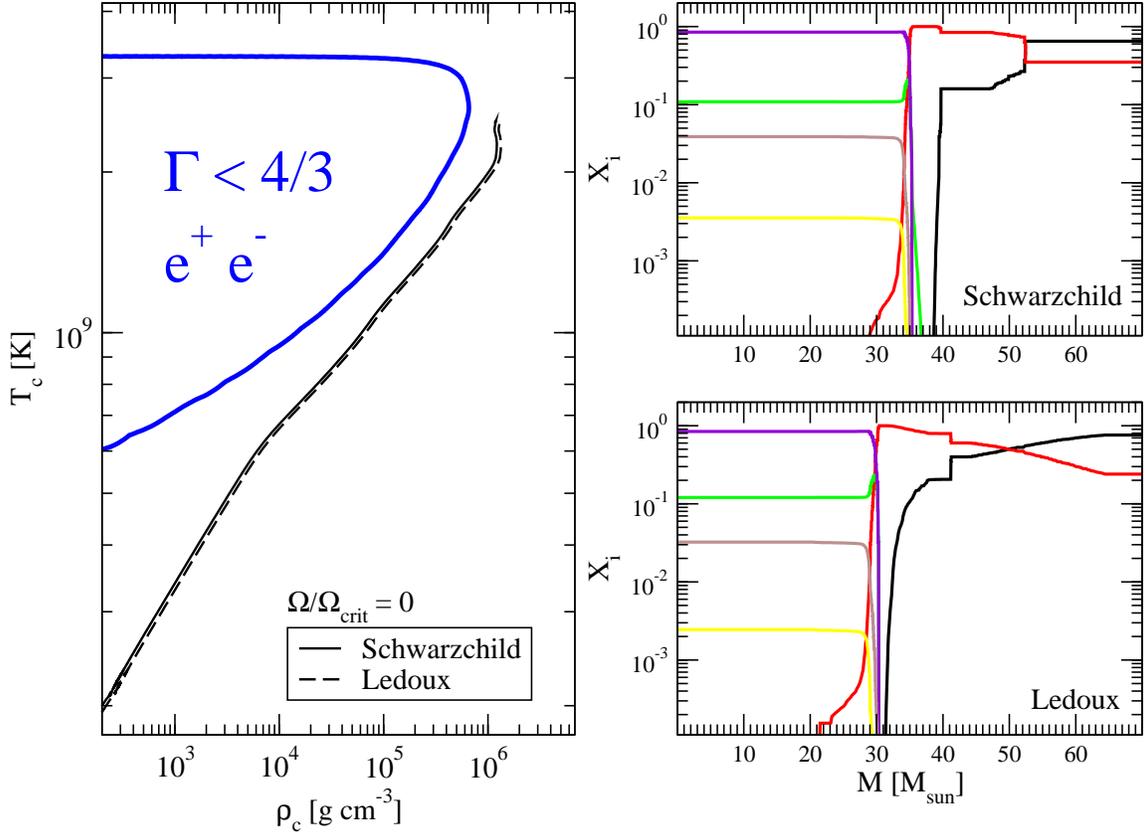}
\caption{{\it Left panel}: Central density and temperature
evolution of the 70~$M_{\odot}$, $\Omega/\Omega_{crit} =$~0 models implementing
the Schwarzschild (solid black curve) and Ledoux (dashed black curve) 
mixing criteria. The solid blue curve 
marks the electron-positron pair instability region where the adiabatic
index is $\Gamma_{1} < 4/3$. 
{\it Upper right panel}: Chemical composition of the 70~$M_{\odot}$, 
$\Omega/\Omega_{crit} =$~0 model
with the Schwarzschild mixing criterion implemented at the time just prior to core oxygen ignition.
{\it Lower right panel}: Chemical composition of the 70~$M_{\odot}$, 
$\Omega/\Omega_{crit} =$~0 model
with the Ledoux mixing criterion implemented at the time just prior to core oxygen ignition.
The specific elements plotted are given in the inset in the upper left panel of Figure 3.}
\end{center}
\end{figure}

\begin{figure}
\begin{center}
\includegraphics[angle=-90,width=18cm]{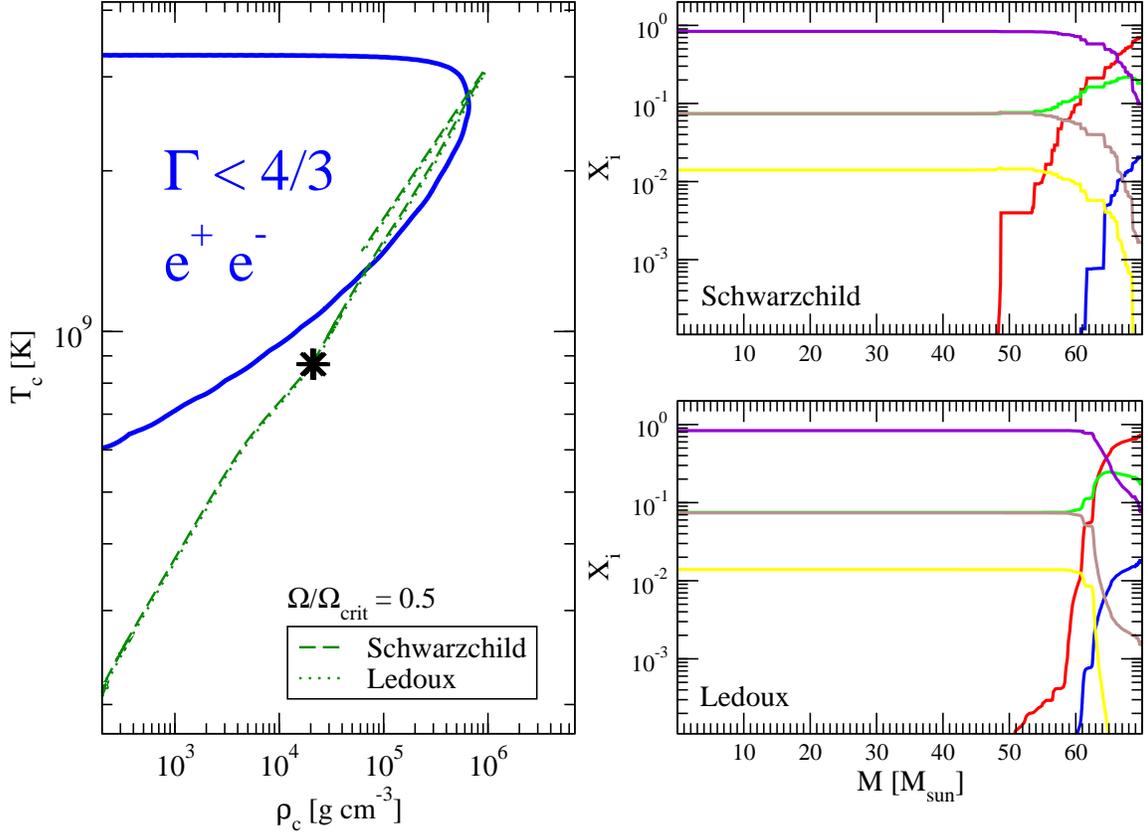}
\caption{{\it Left panel}: Central density and temperature
evolution of the 70~$M_{\odot}$, $\Omega/\Omega_{crit} =$~0.5 models implementing the
Schwarzschild (dashed green curve) and Ledoux (dotted green curve) 
mixing criteria. The solid blue curve 
marks the electron-positron pair instability region where the adiabatic
index is $\Gamma_{1} < 4/3$.  
The black stars mark the point where the models were mapped to
the hydrodynamics code. 
{\it Upper right panel}: Chemical composition of the 70~$M_{\odot}$, 
$\Omega/\Omega_{crit} =$~0.5 model
with the Schwarzschild mixing criterion implemented at the time just prior to core oxygen ignition.
{\it Lower right panel}: Chemical composition of the 70~$M_{\odot}$, 
$\Omega/\Omega_{crit} =$~0.5 model
with the Ledoux mixing criterion implemented at the time just prior to core oxygen ignition.
The specific elements plotted are given in the inset in the upper left panel of Figure 3.}
\end{center}
\end{figure}

\begin{figure}
\begin{center}
\includegraphics[angle=-90,width=18cm]{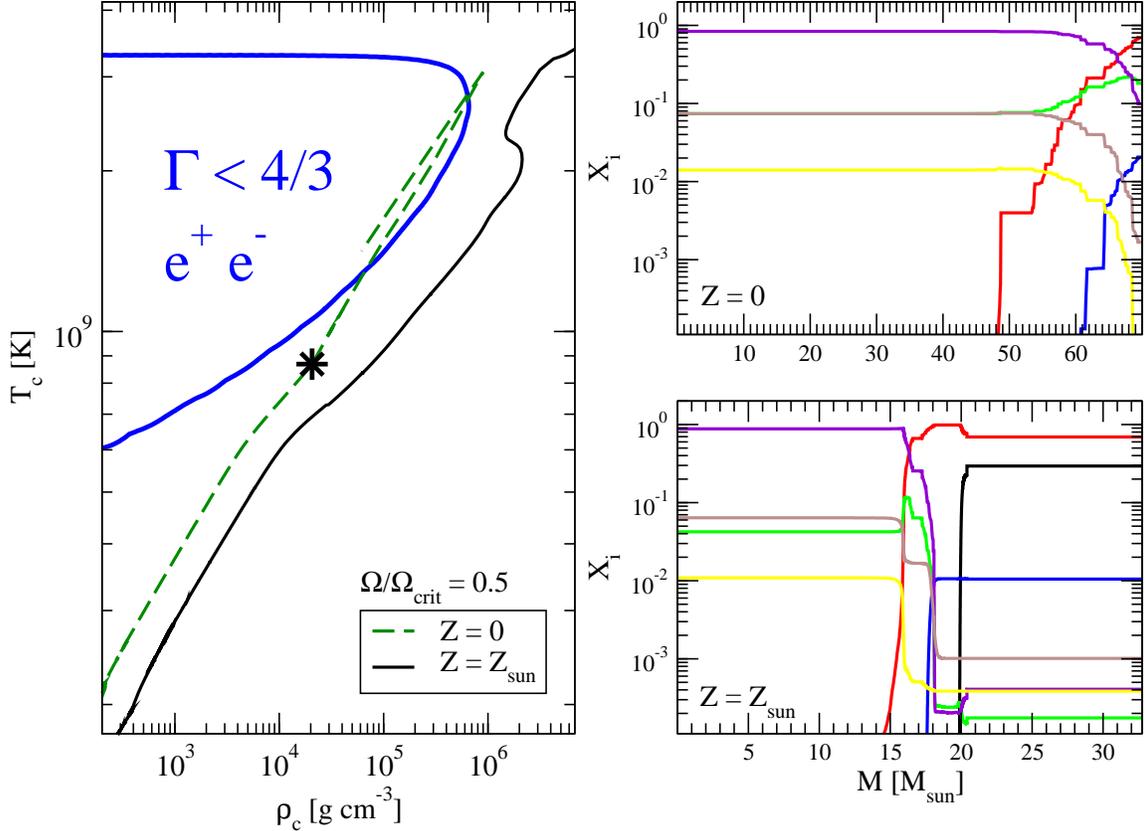}
\caption{{\it Left panel}: Central density and temperature
evolution of the 70~$M_{\odot}$, $\Omega/\Omega_{crit} =$~0.5 models
for $Z =$~0 (dashed green curve) and $Z =$~$Z_{\odot}$ (solid black curve) using the Schwarzschild criterion
and the de Jager, Nieuwenhuijzen \& van der Hucht (1988) mass-loss prescriptions.
The solid blue curve 
marks the electron-positron pair instability region where the adiabatic
index is $\Gamma_{1} < 4/3$.  
The black stars mark the point where the models were mapped to
the hydrodynamics code. 
{\it Upper right panel}: Chemical composition of the 70~$M_{\odot}$ for $Z =$~0 at the time before 
core oxygen ignition.
{\it Lower right panel}: Chemical composition of the 70~$M_{\odot}$ for $Z =$~$Z_{\odot}$
at the time before core oxygen ignition.
The specific elements plotted are given in the inset in the upper left panel of Figure 3.}
\end{center}
\end{figure}

%%%%%%%%%%%%%%TABLE%%%%%%%%%%%%%%%%%%%%%%%%%%%%%%%%%%%%%%%%%%%%%%%%%%%%%%%

\clearpage
%\begin{landscape}
\setcounter{table}{0}
\begin{deluxetable}{lllllllllllllllllllllllllllllllllllllllllllllllccccccc}
\tabletypesize{\tiny}
\tablewidth{0pt}
\tablecaption{Characteristics of the stellar models discussed in this work.}
\tablehead{
\colhead {$M_{ZAMS}$~($M_{\odot}$)} &
\colhead {$\Omega/\Omega_{crit,ZAMS}$} &
\colhead {$T_{9,max}$~$^{a}$} &
\colhead {$\rho_{5,max}$~$^{b}$} &
\colhead {$M_{O-core}$~($M_{\odot}$)} &
\colhead {$X_{^{14}N,surf}$~$^{c}$~($10^{-3}$)} &
\colhead {Fate$^{d}$} \\
}
\startdata
%z$\ast$&0.143&0.143&0.143&0.57&0.57&0.57\\
35  & 0.8 & -    & -    & 35 & 4.70 & CC     \\
\hline
40  & 0.0 & -    & -    & 14 & 0.00 & CC     \\
40  & 0.3 & -    & -    & 25 & 0.43 & CC     \\
40  & 0.5 & -    & -    & 38 &22.64 & CC     \\
40  & 0.8 & 1.16 & 1.03 & 40 & 1.48 & PPISN  \\
\hline
45  & 0.0 & -    & -    & 17 & 0.00 & CC     \\ 
45  & 0.3 & -    & -    & 30 & 1.31 & CC     \\
45  & 0.5 & 1.21 & 0.63 & 44 &15.54 & PPISN  \\
45  & 0.8 & 0.90 & 0.27 & 45 & 5.49 & PPISN  \\
\hline
50  & 0.0 & -    & -    & 19 & 0.00 & CC     \\
50  & 0.3 & -    & -    & 39 & 0.18 & CC     \\
50  & 0.5 & 3.19 & 9.25 & 48 &20.01 & PPISN  \\ 
50  & 0.8 & 1.17 & 1.85 & 50 & 1.32 & PPISN  \\
\hline
55  & 0.0 & -    & -    & 22 & 0.00 & CC     \\
55  & 0.3 & 2.04 & 3.53 & 41 & 0.15 & PPISN  \\
55  & 0.5 & 3.22 & 9.57 & 53 &14.08 & PPISN  \\
55  & 0.8 & 1.08 & 0.88 & 55 & 1.66 & PPISN  \\
\hline
60  & 0.0 & -    & -    & 30 & 0.00 &  CC    \\ 
60  & 0.3 & 2.09 & 3.31 & 45 & 0.04 &  PPISN \\
60  & 0.5 & 2.91 & 8.09 & 57 &26.90 &  PPISN \\
60  & 0.8 & 1.05 & 0.53 & 60 &15.28 &  PPISN \\
\hline
65  & 0.0 & -    & -    & 32 & 0.00 &  CC    \\
65  & 0.3 & 2.37 & 5.37 & 46 & 0.00 &  PPISN \\
65  & 0.5 & 3.49 &12.16 & 60 &26.51 &  PPISN \\ 
65  & 0.8 & 4.45 &35.6  & 65 & 0.00 &  PISN  \\
\hline
70  & 0.0 & -    & -    & 35 & 0.00 &  CC    \\
70$^{\star}$     &0.0   & -  & -    &  30    &  0.00&  CC    \\
70  & 0.3 & 2.58 & 6.90 & 48 & 0.00 &  PPISN \\
70  & 0.5 & 3.03 & 9.22 & 66 &20.58 &  PPISN \\
70$^{\star}$     & 0.5  &3.02& 9.20 &   65   &18.12 &  PPISN \\
70$^{\dagger}$   & 0.5  & -  & -    &   16   &10.51 &  CC    \\
70  & 0.8 & 5.08 &43.18 & 70 & 5.30 &  PISN  \\
\hline
75  & 0.0 & -    & -    & 36 & 0.00 &  CC    \\ 
75  & 0.3 & 3.21 &10.40 & 54 & 0.02 &  PPISN \\
75  & 0.5 & 3.61 &20.00 & 67 &16.68 &  PISN  \\
75  & 0.8 & 5.28 &61.16 & 75 & 0.00 &  PISN  \\
\hline
80  & 0.0 & 1.87 & 2.44 & 40 & 0.00 &  PPISN \\
80  & 0.3 & 1.99 & 2.30 & 59 & 0.00 &  PPISN \\
80  & 0.5 & 4.31 &20.36 & 77 &26.41 &  PISN  \\ 
80  & 0.8 & 2.97 & 8.99 & 80 & 0.00 &  PISN  \\
\hline
85  & 0.0 & 2.09 & 3.61 & 40 & 0.00 &  PPISN \\
85  & 0.3 & 1.53 & 0.81 & 65 & 0.39 &  PISN  \\
85  & 0.5 & 3.46 &10.41 & 79 &24.93 &  PISN  \\
85  & 0.8 & 3.25 & 9.84 & 85 & 8.64 &  PISN  \\
\hline
90  & 0.0 & 1.86 & 1.76 & 45 & 0.00 &  PPISN \\ 
90  & 0.3 & 3.98 &17.44 & 85 &14.24 &  PISN  \\
\hline
95  & 0.0 & 2.75 & 6.59 & 50 & 0.00 &  PPISN \\  
95  & 0.3 & 4.62 &26.97 & 90 &33.25 &  PISN  \\ 
\hline  
110 & 0.0 & 2.46 & 5.85 & 56 & 0.00 &  PPISN \\  
\hline
200 & 0.0 & 5.02 &34.40 &120 & 0.00 &  PISN  \\       
\enddata 
\tablecomments{$^{a}$ In units of $10^{9}$~K. 
$^{b}$ In units of $10^{5}$~g~cm${-3}$.
$^{c}$ We adopt mass fraction of 0.00 to be anything less than $10^{-6}$.
$^{d}$ CC=Core Collapse, PPISN=Pulsational Pair Instability Supernova, PISN=Pair Instability Supernova.
$^{\star}$ With Ledoux criterion for mixing implemented.
$^{\dagger}$ For $Z =$~$Z_{\odot}$ and mass loss.}
\end{deluxetable}

\end{document}